\documentclass[english,reprint,tightenlines,eqsecnum,aps,prd,nofootinbib]{revtex4}
\usepackage[T1]{fontenc}
\usepackage[latin9]{inputenc}
\usepackage{amsmath}
\usepackage{amssymb}
\usepackage{stmaryrd}
\usepackage{graphicx}
\usepackage{esint}

\makeatletter
\@ifundefined{textcolor}{}
{%
 \definecolor{BLACK}{gray}{0}
 \definecolor{WHITE}{gray}{1}
 \definecolor{RED}{rgb}{1,0,0}
 \definecolor{GREEN}{rgb}{0,1,0}
 \definecolor{BLUE}{rgb}{0,0,1}
 \definecolor{CYAN}{cmyk}{1,0,0,0}
 \definecolor{MAGENTA}{cmyk}{0,1,0,0}
 \definecolor{YELLOW}{cmyk}{0,0,1,0}
}

\@ifundefined{showcaptionsetup}{}{%
 \PassOptionsToPackage{caption=false}{subfig}}
\usepackage{subfig}
\makeatother

\usepackage{babel}
\begin{document}

\title{Mode-sum renormalization of $\langle\Phi^{2}\rangle$ for a quantum
scalar field inside a Schwarzschild black hole}

\author{Assaf Lanir, Adam Levi and Amos Ori }

\address{Department of physics, Technion-Israel Institute of Technology, Haifa
32000, Israel}
\begin{abstract}
The full computation of the renormalized expectation values $\langle\Phi^{2}\rangle_{ren}$
and $\langle\hat{T}_{\mu\nu}\rangle_{ren}$ in 4D black hole interiors
has been a long standing challenge, which has impeded the investigation
of quantum effects on the internal structure of black holes for decades.
Employing a recently developed mode sum renormalization scheme to
numerically implement the point-splitting method, we report here the
first computation of $\langle\Phi^{2}\rangle_{ren}$ in Unruh state
in the region inside the event horizon of a 4D Schwarzschild black
hole. We further present its Hartle-Hawking counterpart, which we
calculated using the same method, and obtain a fairly good agreement
with previous results attained using an entirely different method
by Candelas and Jensen in 1986. Our results further agree upon approaching
the event horizon when compared with previous results calculated outside
the black hole. Finally, the results we obtained for Hartle-Hawking
state at the event horizon agree with previous analytical results
published by Candelas in 1980. This work sets the stage for further
explorations of $\langle\Phi^{2}\rangle_{ren}$ and $\langle\hat{T}_{\mu\nu}\rangle_{ren}$
in 4D black hole interiors.
\end{abstract}
\maketitle

\section{Introduction}

It is well known that classical matter fields on black hole (BH) backgrounds
can considerably modify the interior geometry. Consider, for instance,
the unperturbed Reissner-Nordström (RN) and Kerr solutions. Both solutions
possess an inner horizon, which is a perfectly regular null hypersurface.
However, classical perturbations, both linear and nonlinear, were
shown to result in the formation of a weak null curvature singularity
along the ingoing section of the inner horizon, in four dimensional
spinning \cite{perturbations in kerr,perturbations in kerr-3,perturbations in kerr-1,perturbations in kerr-2,Dafermos}
and spherically-symmetric charged BHs \cite{Hiscock evolution,Poisson and Israel,Amos mass inflation,Brady and Smith RN,Burko RN}.
A different example for the modification that classical perturbations
impose upon BH interior geometry is the effective shock wave singularity
developing along the outgoing section of the inner horizon \cite{Shock wave 1,Shock wave 2,Shock wave 3}.
It is therefore reasonable to expect that quantum matter fields would
also affect the BH interior geometry. Indeed, semiclassical general
relativity, in its turn, proved to have the potential to drastically
influence the evolution of BHs. It implies, for example, that BHs
emit radiation and evaporate, as was shown in an analysis by Hawking
\cite{Hawking}. This process results in a spacetime structure radically
different from the classical BH structure. It therefore seems conceivable
that semiclassical stress-energy fluxes could potentially affect the
inner horizon of the RN and Kerr solutions in a different manner than
the classical perturbations considered thus far. In particular, one
may consider the scenario where semiclassical effects convert the
classical weak null singularity into a strong (i.e. tidally destructive)
spacelike one. Or alternatively, one may entertain the thought that
semiclassical effects might actually resolve the strong spacelike
singularity in Schwarzschild spacetime. These are profound issues
that remain as yet unresolved.

Interested as we are in the internal structure of BHs, we must therefore
investigate the semiclassical picture of BH interiors. Semiclassical
gravity considers quantum field theory in curved spacetime, where
gravitation is treated classically, i.e. describing spacetime structure
as a Lorentzian manifold that is equipped with a metric $g_{\mu\nu}$.
At the same time, the matter fields propagating in that classical
background are quantum fields. The relation between spacetime geometry
and the stress-energy of the quantum matter fields is described by
the semiclassical Einstein's field equation
\begin{equation}
G_{\mu\nu}=8\pi\langle\hat{T}_{\mu\nu}\rangle_{ren}.\label{semiclassical Einstein}
\end{equation}
Here $G_{\mu\nu}$ is the Einstein tensor of spacetime, and $\langle\hat{T}_{\mu\nu}\rangle_{ren}$
is the \emph{renormalized stress-energy tensor} (RSET), which is the
renormalized expectation value of the stress-energy tensor operator
$\hat{T}_{\mu\nu}$, associated with the quantum fields. In Eq. \eqref{semiclassical Einstein}
and throughout this paper we adopt standard geometric units $c=G=1$,
along with the metric signature $\left(-+++\right)$.

As already mentioned, we are generally interested in the internal
structure of BHs within the semiclassical framework, and especially
in the effect of quantum fields on the inner horizons of RN and Kerr
BHs. To this end, we have initiated a research program aimed to compute
the RSET on BH backgrounds. In the present work we focus attention
on the interior of a Schwarzschild BH, which is interesting in its
own right. As an example of a quantum field we take for simplicity
a massless scalar field, satisfying the massless Klein-Gordon equation\footnote{The Klein-Gordon equation for a nonminimally-coupled massless scalar
field also includes the term $\xi R\hat{\Phi}$, where $R$ is the
Ricci scalar. However, we are considering here a Schwarzschild spacetime
where the Ricci scalar vanishes. }
\begin{equation}
\boxempty_{g}\hat{\Phi}=0,\label{KG}
\end{equation}
where $\hat{\Phi}$ is the scalar field operator, and the covariant
D'Alembertian operator is used with respect to the background spacetime
metric, denoted by $g$, the Schwarzschild metric in the present case.
Although the physically more interesting object is the RSET, it is
useful to first compute $\langle\Phi^{2}\rangle_{ren}$ as it is technically
simpler, yet it still captures many of the RSET's essential features,
thereby serving as a simple toy-model for it. 

However, the computation of $\langle\Phi^{2}\rangle_{ren}$ (and of
other composite operators, including the RSET) in a general curved
spacetime is a tremendously difficult task. The naive computation
yields a divergent series of modes which requires renormalization.
A renormalization method based on point-splitting was developed by
Christensen \cite{Christensen,WKB outside 1}. He employed the DeWitt-Schwinger
expansion of Feynman's Green function \cite{Dewitt - Dynamical theory of groups and fields,Schwinger}
(see also \cite{Birrell =000026 Davies book}) in order to compute
$\langle\Phi^{2}\rangle_{ren}$ (and similarly the RSET, $\langle\hat{T}_{\mu\nu}\rangle_{ren}$).
The basic idea is to regularize the expectation value by splitting
the point $x$, at which $\langle\Phi^{2}\left(x\right)\rangle_{ren}$
is to be evaluated, into a separated pair of points $x$ and $x'$,
and consider the two-point function $\langle\hat{\Phi}\left(x\right)\hat{\Phi}\left(x'\right)\rangle$.
It is convenient and common - and we shall be adhering to that convention
- to consider, instead of the two-point function, the Hadamard function
which is related in a simple way to the two-point function and is
defined as
\begin{equation}
G^{\left(1\right)}\left(x,x'\right)=\langle\left\{ \hat{\Phi}\left(x\right),\hat{\Phi}\left(x'\right)\right\} \rangle,\label{Hadamard}
\end{equation}
where $\left\{ ,\right\} $ denotes anti-commutation. A specific counter-term,
which is a local geomtric object depending on the background metric
and independent of the quantum state, is then subtracted from the
Hadamard function. Finally the limit $x'\rightarrow x$ is taken,
yielding $\langle\Phi^{2}\rangle_{ren}$. A similar procedure can
be applied to the computation of the RSET and of other composite operators
which are quadratic in the field operator and its derivatives.

This point-splitting scheme works fine in situations where the scalar
field modes can be analytically computed. Unfortunately, in BH backgrounds
the modes must be numerically computed, and as a result the procedure
by Christensen, as is, is impractical. Nevertheless, practical methods
that numerically implement the point-splitting scheme were later developed
by Candelas, Howard, Anderson and others \cite{WKB outside 1,WKB outside 2,WKB outside 3,WKB outside 4,Taylor},
allowing for the computation of $\langle\Phi^{2}\rangle_{ren}$ and
the RSET, usually requiring a high-order WKB expansion for the field
modes. Seeing as this is highly difficult to carry out in the Lorentzian
section of spacetime, a Wick rotation to the Euclidean section is
usually performed. This approach restricts the background spacetimes
where the renormalization procedure can be applied, since a generic
spacetime does not admit a Euclidean section. 

The \textit{pragmatic mode-sum regularization} (PMR) scheme, recently
developed by two of the authors (AL and AO), offers a more versatile
method to numerically implement the point-splitting scheme. It relies
neither on a Euclidean section nor on a WKB expansion. It only requires
a single Killing field in the background spacetime. The PMR method
was used \cite{Adam1,Adam2,Adam3,Adam4,Adam5} to compute $\langle\Phi^{2}\rangle_{ren}$
and the RSET outside the event horizons of the three canonical BH
solutions, namely Schwarzschild, RN (unpublished), and Kerr.

So far, most of the calculations of $\langle\Phi^{2}\rangle_{ren}$
and the RSET were carried out in the regions exterior to the event
horizon of BHs. We are only aware of a single exception presented
in Ref. \cite{Candelas_and_Jensen}, where $\langle\Phi^{2}\rangle_{ren}$
was calculated in the interior region of a Schwarzschild BH, in the
Hartle-Hawking state \cite{HH,Israel on HH}. A possible reason for
that may be that the computation in BH interiors entails intricate
analytic derivations, expressing the Hadamard function by the standard
Eddington-Finkelstein modes, and it further involves tedious numerical
computations in both regions, interior and exterior. The present work
implements the angular-splitting (or ``$\theta$-splitting'') variant
of the PMR method, introduced in \cite{Adam2}, to compute $\langle\Phi^{2}\rangle_{ren}$
in the interior region of a Schwarzschild BH, as a first stage before
computing it in the interior of RN. We employed the PMR method in
the Hartle-Hawking state to recover the results appearing in \cite{Candelas_and_Jensen}
in the domain $M\leq r<2M$, and we found a fairly good agreement.
We further compute $\langle\Phi^{2}\rangle_{ren}$ in the Unruh state
\cite{Unruh} (in the same range of $r$), which is physically more
interesting as it describes an evaporating BH. Our results for $\langle\Phi^{2}\rangle_{ren}$
in both quantum states (i.e. Hartle-Hawking and Unruh) show nice agreement
with previous analogous results obtained outisde the BH \cite{Adam2}
when compared at the event horizon, where both sets of results meet.

The organization of this paper is as follows. In section \ref{sec:Preliminaries}
we introduce necessary preliminaries needed for the computation of
$\langle\Phi^{2}\rangle_{ren}$. Then, in section \ref{PMR } we briefly
present the key steps in the renormalization procedure we use in the
computation. Section \ref{numerical implementation} presents the
details of the numerical implementation and the results we obtained
for $\langle\Phi^{2}\rangle_{ren}$. In section \ref{Discussion}
we discuss our results and possible extensions. 

\section{Preliminaries\label{sec:Preliminaries}}

Before we delve into the details of the numerical implementation of
the angular-splitting variant of the PMR method inside the Schwarzschild
event horizon, let us precede by establishing the basic definitions
of coordinate systems, sets of modes, and quantum states which we
use in the present work. We also write down the form of the Hadamard
function inside the event horizon of a Schwarzschild BH, for both
Hartle-Hawking and Unruh states. This preliminary section summarizes
the relevant definitions and results of \cite{group paper TPF}, with
the reservation that Ref. \cite{group paper TPF} considers RN BHs.
Specializing to the Schwarzschild case simply amounts to taking the
limit where the electric charge of the BH vanishes.

\subsection{Coordinate systems}

The line element of the Schwarzschild solution in the standard Schwarzschild
coordinates takes the form 
\begin{equation}
ds^{2}=-fdt^{2}+f^{-1}dr^{2}+r^{2}\left(d\theta^{2}+\sin^{2}\theta d\varphi^{2}\right),\label{Schwarz. interval}
\end{equation}
where
\begin{equation}
f=1-\frac{2M}{r}.\label{metric f}
\end{equation}
The event horizon, $r=r_{s}$, is located at the root of $f$, i.e.
\[
r_{s}=2M.
\]
The surface gravity parameter, $\kappa$, is given by
\[
\kappa=\frac{1}{4M}.
\]

We define the tortoise coordinate, $r_{*}$, in both the interior
and the exterior regions, using the standard relation
\[
\frac{dr}{dr_{*}}=1-\frac{2M}{r}.
\]
Specifically, we choose the integration constants in the interior
and the exterior regions such that in both regions 
\[
r_{*}=r+2M\,\ln\left(\left|1-\frac{r}{2M}\right|\right).
\]
Note that $r_{s}$ corresponds to $r_{*}\rightarrow-\infty$ (both
inside and outside the BH), and $r=0$ to $r_{*}\rightarrow0$. 

\begin{figure}
\begin{centering}
\includegraphics{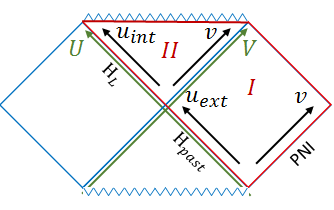}
\par\end{centering}
\caption{Penrose diagram of Schwarzschild spacetime. In the exterior region
(region $I$), we use the external Eddington-Finkelstein coordinates,
while in the interior (region $II$), we use the internal ones. In
addition, the Kruskal coordinate system is shown in green and is defined
throughout both regions $I$ and $II$. The red-framed area denotes
the region in the eternal Schwarzschild spacetime which concerns this
paper, i.e. regions $I$ and $II$. \label{fig:Penrose diag}}
\end{figure}

The Eddington-Finkelstein coordinates are defined in the exterior
and interior regions by
\[
u_{\mathrm{ext}}=t-r_{*}\quad,\quad v=t+r_{*},\quad\left(\textrm{outside}\right)
\]
and
\[
u_{\mathrm{int}}=r_{*}-t\quad,\quad v=r_{*}+t,\quad\left(\textrm{inside}\right).
\]
Note that the $v$ coordinate is continuously defined throughout both
regions I and II of Fig. \ref{fig:Penrose diag}. 

The Kruskal coordinates are defined in terms of the exterior and interior
Eddington-Finkelstein coordinates by
\begin{equation}
U\left(u_{\mathrm{ext}}\right)=-\frac{1}{\kappa}\exp\left(-\kappa u_{\mathrm{ext}}\right)\quad,\quad V\left(v\right)=\frac{1}{\kappa}\exp\left(\kappa v\right),\quad\left(\textrm{outside}\right)\label{Kruskal ext}
\end{equation}
and
\begin{equation}
U\left(u_{\mathrm{int}}\right)=\frac{1}{\kappa}\exp\left(\kappa u_{\mathrm{int}}\right)\quad,\quad V\left(v\right)=\frac{1}{\kappa}\exp\left(\kappa v\right),\quad\left(\textrm{inside}\right).\label{Kruskal int}
\end{equation}

We make the following notations: $H_{\mathrm{past}}$ denotes the
past horizon {[}i.e. the hypersurface $\left(U<0\,,\,V=0\right)${]},
and PNI denotes past null infinity {[}i.e. $\left(U=-\infty\,,\,V>0\right)${]}.
$H_{L}$ is the ``left event horizon'' $\left(U>0\,,\,V=0\right)$,
and $H_{R}$ is the ``right event horizon'' $\left(U=0\,,\,V>0\right)$.
See Fig. \ref{fig:Penrose diag}.

\subsection{Sets of modes and quantum states\label{Modes and states}}

Recall that the matter field we are considering here is a massless
quantum scalar field satisfying the D'Alembertian equation \eqref{KG}
on Schwarzschild background. This field can be decomposed into sets
of modes satisfying Eq. \eqref{KG}. By choosing certain sets of modes,
one can define several quantum states of interest. To this end, it
will be useful to consider various complete sets of modes outside
and inside the event horizon.

Let us begin with the definition of the \emph{Unruh modes} for positive
$\omega$, which we deonte by $g_{\omega lm}^{\mathrm{up}}$ and $g_{\omega lm}^{\mathrm{in}}$
. We first use the spherical symmetry to decompose the modes in the
following standard way: 
\begin{equation}
g_{\omega lm}^{\Lambda}\left(x\right)=\omega^{-1/2}C_{lm}\left(x\right)\tilde{g}_{\omega l}^{\Lambda}\left(x\right),\label{full Unruh modes}
\end{equation}
where 
\begin{equation}
C_{lm}\left(x\right)=\left(4\pi\right)^{-1/2}\frac{1}{r}Y_{lm}\left(\theta,\varphi\right).\label{C}
\end{equation}
Here the index $\varLambda$ denotes ``in'' and ``up'', and the
factor $1/\sqrt{4\pi\omega}$ was introduced to ensure proper normalization
under the Klein-Gordon inner product. Furthermore, $\tilde{g}_{\omega l}^{\Lambda}$
are solutions of the following two-dimensional wave equation, obtained
by substituting Eq. \eqref{full Unruh modes} in Eq. \eqref{KG}:
\begin{equation}
\tilde{g}_{,r_{*}r_{*}}^{\Lambda}-\tilde{g}_{,tt}^{\Lambda}=V_{l}\left(r\right)\tilde{g}^{\Lambda},\label{wave eq}
\end{equation}
where
\begin{equation}
V_{l}\left(r\right)=\left(1-\frac{2M}{r}\right)\left[\frac{l\left(l+1\right)}{r^{2}}+\frac{2M}{r^{3}}\right].\label{Potential}
\end{equation}
The Unruh modes are then defined by requiring the two sets of independent
solutions, $\tilde{g}_{\omega l}^{\mathrm{up}}$ and $\tilde{g}_{\omega l}^{\mathrm{in}}$,
to satisfy the following initial conditions: 
\begin{equation}
\tilde{g}_{\omega l}^{\mathrm{up}}=\left\{ \begin{array}{c}
e^{-i\omega U}\quad,\quad H_{\mathrm{past}}\\
e^{-i\omega U}\quad,\quad H_{L}\\
0\quad,\quad\mathrm{\mathrm{PNI}}
\end{array}\right.\qquad,\qquad\tilde{g}_{\omega l}^{\mathrm{in}}=\left\{ \begin{array}{c}
0\quad,\quad H_{\mathrm{past}}\\
0\quad,\quad H_{L}\\
e^{-i\omega v}\quad,\quad\mathrm{\mathrm{PNI}}
\end{array}\right..\label{g-in-up}
\end{equation}
Note that the Unruh modes are defined both inside and outside the
event horizon, i.e. they are defined throughout the red-framed regions
$I$ and $II$ in Fig. \ref{fig:Penrose diag}. 

Using these modes, the quantum scalar field operator is decomposed
as follows:
\begin{equation}
\hat{\Phi}\left(x\right)=\intop_{0}^{\infty}d\omega\sum_{\Lambda,l,m}\left[g_{\omega lm}^{\Lambda}\left(x\right)\hat{a}_{\omega lm}^{\Lambda}+g_{\omega lm}^{\Lambda*}\left(x\right)\hat{a}_{\omega lm}^{\Lambda\dagger}\right],\label{operator decomp. Unruh}
\end{equation}
and the Unruh state, $\left|0\right\rangle _{U}$, is defined as the
quantum state annihilated by the operators $\hat{a}_{\omega lm}^{\Lambda}$
appearing in \eqref{operator decomp. Unruh}, i.e.
\[
\hat{a}_{\omega lm}^{\Lambda}\left|0\right\rangle _{U}=0
\]
(for all $\Lambda$ and $\omega lm$). As was mentioned in the introduction,
the Unruh state is physically interesting, as it describes an evaporation
of a BH \cite{Unruh}. The vacuum expectation value of the stress-energy
tensor in this state is regular at the event horizon ($H_{R}$), but
not at the past horizon ($H_{\mathrm{past}}$) \cite{Candelas}.

Let us now procede to the definition of the \emph{outer Eddington-Finkelstein
modes}, $f_{\omega lm}^{\mathrm{up}}$ and $f_{\omega lm}^{\mathrm{in}}$,
for positive $\omega$ (these are also known as the \emph{Boulware
modes}, as they are closely related to the Boulware state \cite{Boulware}).
Again, we decompose the modes like we did in Eq. \eqref{full Unruh modes}
above:
\begin{equation}
f_{\omega lm}^{\Lambda}\left(x\right)=\left|\omega\right|^{-1/2}C_{lm}\left(x\right)\tilde{f}_{\omega l}^{\Lambda}\left(x\right),\label{full Edding modes}
\end{equation}
where $C_{lm}$ is the same function appearing in Eq. \eqref{C} and
the index $\Lambda$ again stands for ``in'' and ``up''. The functions
$\tilde{f}_{\omega l}^{\Lambda}$, like $\tilde{g}_{\omega l}^{\Lambda}$,
also satisfy the wave equation \eqref{wave eq}, but with the significant
difference that they are only defined in region $I$ (see Fig. \ref{fig:Penrose diag}).
To complement the definition of these functions we require the following
initial conditions: 

\begin{equation}
\tilde{f}_{\omega l}^{\mathrm{in}}=\left\{ \begin{array}{c}
0\quad,\quad H_{\mathrm{past}}\\
e^{-i\omega v}\quad,\quad\mathrm{PNI}
\end{array}\right.\qquad,\qquad\tilde{f}_{\omega l}^{\mathrm{up}}=\left\{ \begin{array}{c}
e^{-i\omega u_{\mathrm{ext}}}\quad,\quad H_{\mathrm{past}}\\
0\quad,\quad\mathrm{PNI}
\end{array}\right..\label{initial conditions in and up}
\end{equation}

Due to the staticity of the background metric, the outer Eddington-Finkelstein\emph{
}modes $\tilde{f}_{\omega l}^{\mathrm{in}}$ and $\tilde{f}_{\omega l}^{\mathrm{up}}$
can be further decomposed into a time-dependent part and a radial
function, satisfying an \emph{ordinary} differential equation (ODE),
which can be readily solved numerically. The decomposition is as follows:
\begin{equation}
\tilde{f}_{\omega l}^{\mathrm{in}}\left(r,t\right)=e^{-i\omega t}\varPsi_{\omega l}^{\mathrm{in}}\left(r\right)\,\,,\,\,\tilde{f}_{\omega l}^{\mathrm{up}}\left(r,t\right)=e^{-i\omega t}\varPsi_{\omega l}^{\mathrm{up}}\left(r\right).\label{decomp rad ext}
\end{equation}
Substituting these decompositions into Eq. \eqref{wave eq} yields
the following radial equation for $\varPsi_{\omega l}^{\mathrm{\Lambda}}$:
\begin{equation}
\varPsi_{\omega l,r_{*}r_{*}}^{\mathrm{\Lambda}}+\left[\omega^{2}-V_{l}\left(r\right)\right]\varPsi_{\omega l}^{\mathrm{\Lambda}}=0,\label{radial-exterior}
\end{equation}
where the effective potential $V_{l}$ is given by Eq. \eqref{Potential}.
In terms of the radial functions $\varPsi_{\omega l}^{\mathrm{\Lambda}}$,
the initial conditions given in Eq. \eqref{initial conditions in and up}
are translated to
\begin{equation}
\varPsi_{\omega l}^{\mathrm{in}}\left(r\right)\cong\left\{ \begin{array}{c}
\tau_{\omega l}^{\mathrm{in}}e^{-i\omega r_{*}}\quad,\quad r_{*}\rightarrow-\infty\\
e^{-i\omega r_{*}}+\rho_{\omega l}^{\mathrm{in}}e^{i\omega r_{*}}\quad,\quad r_{*}\rightarrow\infty
\end{array}\right.\label{bc exterior in}
\end{equation}
and 

\begin{equation}
\varPsi_{\omega l}^{\mathrm{up}}\left(r\right)\cong\left\{ \begin{array}{c}
e^{i\omega r_{*}}+\rho_{\omega l}^{\mathrm{up}}e^{-i\omega r_{*}}\quad,\quad r_{*}\rightarrow-\infty\\
\tau_{\omega l}^{\mathrm{up}}e^{i\omega r_{*}}\quad,\quad r_{*}\rightarrow\infty
\end{array}\right.,\label{bc exterior up}
\end{equation}
where $\rho_{\omega l}^{\Lambda}$ and $\tau_{\omega l}^{\varLambda}$
are the reflection and transmission coefficients (corresponding to
the mode $\tilde{f}_{\omega l}^{\Lambda}$), respectively. Solving
numerically Eq. \eqref{radial-exterior} together with the boundary
conditions \eqref{bc exterior in} and \eqref{bc exterior up} yields
$\varPsi_{\omega l}^{\mathrm{\Lambda}}\left(r\right)$, which then
gives the modes $f_{\omega lm}^{\mathrm{in}}$ and $f_{\omega lm}^{\mathrm{\mathrm{\mathrm{up}}}}$
using Eqs. \eqref{decomp rad ext} and \eqref{full Edding modes}. 

We can now similarly define the two sets of \emph{inner Eddington-Finkelstein
modes}, which are again decomposed according to Eq. \eqref{full Edding modes},
and the corresponding functions $\tilde{f}_{\omega l}^{\Lambda}$
also satisfy Eq. \eqref{wave eq}. It is very analogous to the above
definition of the \emph{outer Eddington-Finkelstein modes}, except
that here $\Lambda$ denotes ``right'' ($R$) and ``left'' ($L$),
corresponding to the following initial conditions on the left and
right event horizons \footnote{As was already stressed in \cite{group paper TPF}, the ``right''
and ``left'' modes defined in the interior region of the BH are
only introduced for mathematical convenience and are not used here
for the definition of any quantum state. }: 

\begin{equation}
\tilde{f}_{\omega l}^{\mathrm{L}}=\left\{ \begin{array}{c}
e^{-i\omega u_{\mathrm{int}}}\quad,\quad H_{L}\\
0\quad,\quad H_{R}
\end{array}\right.\qquad,\qquad\tilde{f}_{\omega l}^{\mathrm{R}}=\left\{ \begin{array}{c}
0\quad,\quad H_{L}\\
e^{-i\omega v}\quad,\quad H_{R}
\end{array}\right..\label{initial conditions L and R}
\end{equation}
Note that these modes are defined only in region $II$, depicted in
Fig. \ref{fig:Penrose diag}.

Just like the outer Eddington-Finkelstein\emph{ }modes, the inner
Eddington-Finkelstein\emph{ }modes $\tilde{f}_{\omega l}^{\mathrm{L}}$
and $\tilde{f}_{\omega l}^{\mathrm{R}}$ can be further decomposed
into a $t$-dependent part and a radial function, satisfying the same
radial equation \eqref{radial-exterior} as the external radial function
$\varPsi_{\omega l}^{\mathrm{\Lambda}}$ (with some, obviously different,
appropriate initial conditions which will be specified below). Note
however the important fact that upon crossing the event horizon from
the exterior region to the interior, the roles of the $t$ and $r$
coordinates as timelike and spacelike, respectively, are reversed.
As a result, oustide the BH there is a single $t$-dependent part,
namely $e^{-i\omega t}$, in the decomposition of both $\tilde{f}_{\omega l}^{\mathrm{in}}$
and $\tilde{f}_{\omega l}^{\mathrm{up}}$. However, inside the BH
the decomposition of $\tilde{f}_{\omega l}^{\mathrm{L}}$ and $\tilde{f}_{\omega l}^{\mathrm{R}}$
is as follows: 
\begin{equation}
\tilde{f}_{\omega l}^{\mathrm{L}}\left(r,t\right)=e^{i\omega t}\psi_{\omega l}\left(r\right)\,\,,\,\,\tilde{f}_{\omega l}^{\mathrm{R}}\left(r,t\right)=e^{-i\omega t}\psi_{\omega l}\left(r\right).\label{decomp rad int}
\end{equation}
Notice that in \eqref{decomp rad int} the two modes differ in their
$t$-dependent part. At the same time, these two modes share a \emph{single}
radial function $\psi_{\omega l}\left(r\right)$ - unlike the external
Eddington-Finkelstein modes which involve two different radial functions,
$\varPsi_{\omega l}^{\mathrm{in}}\left(r\right)$ and $\varPsi_{\omega l}^{\mathrm{up}}\left(r\right)$.
Substitution of the decompositions \eqref{decomp rad int} into Eq.
\eqref{wave eq} yields the aforementioned radial equation for $\psi_{\omega l}$,
namely
\begin{equation}
\psi_{\omega l,r_{*}r_{*}}+\left[\omega^{2}-V_{l}\left(r\right)\right]\psi_{\omega l}=0.\label{radial interior}
\end{equation}
In terms of this radial function, the initial conditions given in
Eq. \eqref{initial conditions L and R} reduce to the single condition
\begin{equation}
\psi_{\omega l}\cong e^{-i\omega r_{*}},\,\,\,r_{*}\rightarrow-\infty.\label{bdry cond}
\end{equation}
Then, solving numerically the ODE \eqref{radial interior} together
with the initial condition \eqref{bdry cond} yields $\psi_{\omega l}\left(r\right)$.
Using Eqs. \eqref{decomp rad int}, \eqref{full Edding modes} and
\eqref{C} will then give us the modes $f_{\omega lm}^{\mathrm{L}}$
and $f_{\omega lm}^{\mathrm{\mathrm{R}}}$. 

Finally, we are also interested in the Hartle-Hawking state \cite{HH,Israel on HH},
denoted by $\left|0\right\rangle _{H}$. Unlike the Boulware state,
the Hartle-Hawking state does not correspond to the conventional concept
of a vacuum. The RSET is regular on the event horizons, both past
($H_{\mathrm{past}}$) and future ($H_{R}$), but the state is not
empty at infinity. In fact, it corresponds to a thermal bath of radiation
at infinity. Although it is customary to define it using an analytic
continuation to the Euclidean section, for our current purposes we
are merely interested in the mode structure of this state; The Hartle-Hawking
modes assume the form of Kruskal modes on both the past ($H_{\mathrm{past}}$)
and furture ($H_{R}$) event horizons, more specifically, $e^{-i\omega U}$
at $H_{\mathrm{past}}$ and $e^{-i\omega V}$ at $H_{R}$. 

\subsection{The Unruh and Hartle-Hawking states Hadamard functions inside the
black hole\label{TPF inside}}

Recall that the inner Eddington-Finkelstein modes can be expressed
in terms of a single radial function, $\psi_{\omega l}\left(r\right)$
{[}see Eq. \eqref{decomp rad int}{]} which can be computed numerically.
We intend to use this fact, and we are thus in need for an expression
of the Hadamard function, inside the BH, in terms of the inner Eddington-Finkelstein
modes (in both states of interest, namely Unruh and Hartle-Hawking). 

Recall that in Eq. \eqref{operator decomp. Unruh} the scalar field
operator was decomposed in terms of the Unruh modes. Substitution
of this expression into the definition of the Hadamard function {[}see
Eq. \eqref{Hadamard}{]}, yields a mode-sum expression for the Hadamard
function in terms of the Unruh modes. As the latter are continuously
defined throughout both regions $I$ and $II$ of Fig. \ref{fig:Penrose diag}
(as discussed in Sec. \ref{Modes and states}), this expression applies
to both the exterior and interior regions of the BH spacetime. As
is thoroughly elaborated in \cite{group paper TPF}, in both the exterior
and interior regions, these Unruh modes can be re-expressed in terms
of the outer and inner Eddington-Finkelstein modes. Using these relations,
expressions for the Unruh state two-point function outside and inside
the BH, in terms of the outer and inner Eddington-Finkelstein modes,
may be obtained. 

This procedure was carried out in \cite{Candelas} for the Feynman
propagator outside the event horizon of Schwarzschild spacetime. When
this procedure is implemented to the Hadamard Green function it yields

\begin{equation}
G_{U}^{\left(1\right)}\left(x,x'\right)=\intop_{0}^{\infty}d\omega\sum_{l,m}\left[\coth\left(\frac{\pi\omega}{\kappa}\right)\left\{ f_{\omega lm}^{\mathrm{up}}\left(x\right),f_{\omega lm}^{\mathrm{up}*}\left(x'\right)\right\} +\left\{ f_{\omega lm}^{\mathrm{in}}\left(x\right),f_{\omega lm}^{\mathrm{in}*}\left(x'\right)\right\} \right],\label{TPF_U outside}
\end{equation}
where the subscript $U$ stands for ``Unruh state'', and the curly
brackets denote symmetrization with respect to the arguments $x$
and $x'$, i.e. 
\[
\left\{ A\left(x\right),B\left(x'\right)\right\} =A\left(x\right)B\left(x'\right)+A\left(x'\right)B\left(x\right).
\]
A similar procedure can be applied to the Hartle-Hawking Hadamard
function, yielding \cite{Candelas}:
\begin{equation}
G_{H}^{\left(1\right)}\left(x,x'\right)=\intop_{0}^{\infty}d\omega\sum_{l,m}\coth\left(\frac{\pi\omega}{\kappa}\right)\left[\left\{ f_{\omega lm}^{\mathrm{up}}\left(x\right),f_{\omega lm}^{\mathrm{up}*}\left(x'\right)\right\} +\left\{ f_{\omega lm}^{\mathrm{in}}\left(x\right),f_{\omega lm}^{\mathrm{in}*}\left(x'\right)\right\} \right],\label{TPF_H outside}
\end{equation}
where the subscript $H$ stands for ``Hartle-Hawking state''. This
expression may actually be obtained by replacing the above Unruh modes
with corresponding modes associated with the Hartle-Hawking state,
which assume the form of Kruskal modes at both the past and future
horizons\footnote{More technically, this may be achieved by changing $e^{-i\omega v}\to e^{-i\omega V}$
at PNI in Eq. \eqref{g-in-up}.}.

However, for the purpose of the present work we are interested in
the expressions for the Unruh state and Hartle-Hawking state Hadamard
functions \emph{inside} the BH. These were obtained in \cite{group paper TPF}
and we merely quote here the final results, which we express here
in the following schematic form: 
\[
G_{U}^{\left(1\right)}\left(x,x'\right)=\intop_{0}^{\infty}d\omega\sum_{l,m}\tilde{E}_{\omega lm}^{U}\left(x,\,x'\right)
\]
and
\[
G_{H}^{\left(1\right)}\left(x,x'\right)=\intop_{0}^{\infty}d\omega\sum_{l,m}\tilde{E}_{\omega lm}^{H}\left(x,\,x'\right),
\]
where
\[
\tilde{E}_{\omega lm}^{U}\left(x,\,x'\right)=\coth\left(\frac{\pi\omega}{\kappa}\right)\left\{ f_{\omega lm}^{\mathrm{L}}\left(x\right),f_{\omega lm}^{\mathrm{L}*}\left(x'\right)\right\} +\left[\coth\left(\frac{\pi\omega}{\kappa}\right)\left|\rho_{\omega l}^{\mathrm{up}}\right|^{2}+\left|\tau_{\omega l}^{\mathrm{up}}\right|^{2}\right]\left\{ f_{\omega lm}^{\mathrm{R}}\left(x\right),f_{\omega lm}^{\mathrm{R}*}\left(x'\right)\right\} 
\]
\begin{equation}
+2\mathrm{csch}\left(\frac{\pi\omega}{\kappa}\right)\mathrm{Re}\left[\rho_{\omega l}^{\mathrm{up}}\left\{ f_{\omega lm}^{\mathrm{R}}\left(x\right),f_{\left(-\omega\right)lm}^{\mathrm{L}*}\left(x'\right)\right\} \right],\label{Hadamard_U inside}
\end{equation}
and
\[
\tilde{E}_{\omega lm}^{H}\left(x,\,x'\right)=\coth\left(\frac{\pi\omega}{\kappa}\right)\left[\left\{ f_{\omega lm}^{\mathrm{L}}\left(x\right),f_{\omega lm}^{\mathrm{L}*}\left(x'\right)\right\} +\left\{ f_{\omega lm}^{\mathrm{R}}\left(x\right),f_{\omega lm}^{\mathrm{R}*}\left(x'\right)\right\} \right]
\]
\begin{equation}
+2\mathrm{csch}\left(\frac{\pi\omega}{\kappa}\right)\mathrm{Re}\left[\rho_{\omega l}^{\mathrm{up}}\left\{ f_{\omega lm}^{\mathrm{R}}\left(x\right),f_{\left(-\omega\right)lm}^{\mathrm{L}*}\left(x'\right)\right\} \right].\label{Hadamard_H inside}
\end{equation}

As already mentioned, the inner Eddington-Finkelstein modes can be
computed {[}via Eqs. \eqref{decomp rad int} and \eqref{full Edding modes}{]}
by numerically solving the ODE \eqref{radial interior} for $\psi_{\omega l}$
in the interior region together with the initial condition given by
Eq. \eqref{bdry cond}. This concludes the numerical computation of
the Unruh and Hartle-Hawking states Hadamard functions in the BH interior. 

\section{Angular-splitting variant of the PMR method: the practical recipe
\label{PMR }}

Recall that the task at hand is to compute the renormalized $\langle\hat{\Phi}^{2}\rangle$
in the Schwarzchild BH interior. As mentioned in the introduction,
a frequently employed technique to regularize $\langle\hat{\Phi}^{2}\left(x\right)\rangle$
is point splitting. This method may be recast in the form
\begin{equation}
\langle\Phi^{2}\left(x\right)\rangle_{ren}=\lim_{x'\rightarrow x}\left[\left\langle \hat{\Phi}\left(x\right)\hat{\Phi}\left(x'\right)\right\rangle -G_{DS}\left(x,x'\right)\right].\label{point split. ren.}
\end{equation}
For our purposes it will be convenient to rewrite Eq. \eqref{point split. ren.}
in terms of the Hadamard function, as follows:
\begin{equation}
\langle\Phi^{2}\left(x\right)\rangle_{ren}=\lim_{x'\rightarrow x}\left[\frac{1}{2}G^{\left(1\right)}\left(x,x'\right)-G_{DS}\left(x,x'\right)\right].\label{point split. ren.-1}
\end{equation}
In the last two equations, $G_{DS}\left(x,x'\right)$ is the DeWitt-Schwinger
counterterm, which in our case, that is for a massless scalar field
propagating in a Schwarzschild background spacetime, takes the form\footnote{This expression for the DeWitt-Schwinger counterterm actually applies
to a massless scalar field propagating in any vacuum background spacetime.} \cite{Christensen,WKB outside 4}
\begin{equation}
G_{DS}\left(x,x'\right)=\frac{1}{8\pi^{2}\sigma}.\label{DS counterterm}
\end{equation}
Here $\sigma$ is the biscalar of geodesic separation, equal to one
half the square of the distance between the points $x$ and $x'$
along the shortest geodesic connecting them. A key feature of the
DeWitt-Schwinger counterterm is that it is a purely local geometric
object, independent of the quantum state, and it fully embodies the
singular part of the Hadamard function at the coincidence limit. 

As we already mentioned in the introduction, the field's modes are
usually not known analytically, and in particular in BH background
spacetimes, the modes can only be numerically computed. In such cases,
the direct evaluation of the coincidence limit of the splitted expression
in Eq. \eqref{point split. ren.-1} becomes impractical, especially
because it requires increasingly high numerical accuracy when $x'$
approaches $x$. This obstacle is overcome by the PMR method. We will
not be reviewing the construction of the PMR method here. For a detailed
exposition of the method, see \cite{Adam1,Adam2}. In what follows
we merely present the key steps involved in the implementation of
the angular-splitting variant \cite{Adam2} of the method.

In Eqs. \eqref{Hadamard_U inside} and \eqref{Hadamard_H inside}
we expressed the Hadamard functions in Unruh and Hartle-Hawking states,
respectively, in terms of the inner Eddington-Finkelstein modes, which
will be our modes of interest in what follows. Recall from Eqs. \eqref{full Edding modes}
and \eqref{decomp rad int} that the field modes can be decomposed
as follows: 
\begin{equation}
f_{\omega lm}^{\mathrm{L}}\left(x\right)=e^{i\omega t}Y_{lm}\left(\theta,\varphi\right)\bar{\psi}_{\omega l}\left(r\right)\,\,,\,\,f_{\omega lm}^{R}\left(x\right)=e^{-i\omega t}Y_{lm}\left(\theta,\varphi\right)\bar{\psi}_{\omega l}\left(r\right).\label{decomp rad int-1}
\end{equation}
where we introduced the notation
\begin{equation}
\bar{\psi}_{\omega l}\left(r\right)=\left(4\pi\left|\omega\right|\right)^{-1/2}\frac{1}{r}\psi_{\omega l}\left(r\right).\label{psi bar}
\end{equation}
The functions $\psi_{\omega l}$ satisfy the ODE \eqref{radial interior},
along with the initial condition given by Eq. \eqref{bdry cond}.

The final outcome of the PMR method is entirely expressed in terms
of the coincidence limit $x'\rightarrow x$ of the quantities $\tilde{E}_{\omega lm}^{U}$
and $\tilde{E}_{\omega lm}^{H}$ defined in Eqs. \eqref{Hadamard_U inside}
and \eqref{Hadamard_H inside} respectively. We therefore define the
coincidence limit of these two quantities (multiplied by $2\pi/\left(2l+1\right)$
for later convenience):

\begin{equation}
E_{\omega l}^{U}\left(r\right)=\sum_{m}\frac{2\pi}{2l+1}\tilde{E}_{\omega lm}^{U}\left(x,\,x'=x\right)\,\,\,\,,\,\,\,\,E_{\omega l}^{H}\left(r\right)=\sum_{m}\frac{2\pi}{2l+1}\tilde{E}_{\omega lm}^{H}\left(x,\,x'=x\right)\label{coinc E 1}
\end{equation}
Note that the dependence on $\theta$, $\varphi$ and $t$ cancels
out upon taking the coincidence limit and summing over $m$. It is
convenient to express $E_{\omega l}^{U}\left(r\right)$ and $E_{\omega l}^{H}\left(r\right)$
directly in terms of the radial function $\psi_{\omega l}$ {[}rather
than the functions $f_{\omega lm}^{\Lambda}$ appearing in Eqs. \eqref{Hadamard_U inside}
and \eqref{Hadamard_H inside}{]}, since this is the function we actually
obtain by solving numerically the radial equation \eqref{radial interior}.
To this end we recall that Eqs. \eqref{decomp rad int-1} and \eqref{psi bar}
provide the relation between the functions $f_{\omega lm}^{\Lambda}$
and $\psi_{\omega l}$, and further note that $\psi_{\left(-\omega\right)l}=\psi_{\omega l}^{*}$.
It thus readily follows that 
\begin{equation}
E_{\omega l}^{U}\left(r\right)=\left[\coth\left(\frac{\pi\omega}{\kappa}\right)\left(\left|\rho_{\omega l}^{\mathrm{up}}\right|^{2}+1\right)+\left|\tau_{\omega l}^{\mathrm{up}}\right|^{2}\right]\left|\bar{\psi}_{\omega l}\right|^{2}+2\mathrm{csch}\left(\frac{\pi\omega}{\kappa}\right)\mathrm{Re}\left(\rho_{\omega l}^{\mathrm{up}}\bar{\psi}_{\omega l}^{2}\right)\label{E_Unruh}
\end{equation}
and 
\begin{equation}
E_{\omega l}^{H}\left(r\right)=2\left[\coth\left(\frac{\pi\omega}{\kappa}\right)\left|\bar{\psi}_{\omega l}\right|^{2}+\mathrm{csch}\left(\frac{\pi\omega}{\kappa}\right)\mathrm{Re}\left(\rho_{\omega l}^{\mathrm{up}}\bar{\psi}_{\omega l}^{2}\right)\right],\label{E_HH}
\end{equation}
where we again used Eq. \eqref{psi bar} to translate these expressions
from $\psi_{\omega l}$ to $\bar{\psi}_{\omega l}$ for compactness. 

For convenience, in what follows unless specifically stated otherwise,
we shall denote both $E_{\omega l}^{U}$ and $E_{\omega l}^{H}$ as
$E_{\omega l}$ because the next stages treat them in exactly the
same way. Following the PMR prescription, we define the following
integral 
\begin{equation}
F\left(l,\,r\right)\equiv\intop_{0}^{\infty}d\omega\left[E_{\omega l}\left(r\right)-E_{\omega,l=0}\left(r\right)\right].\label{F}
\end{equation}
We further define
\begin{equation}
F_{sing}\left(l,\,r\right)=-8\pi a\left(r\right)h\left(l\right),\label{F_singular}
\end{equation}
which captures the singular piece of the function $F\left(l,\,r\right)$.
Here, the function $h\left(l\right)$ is the \emph{Harmonic Number
}defined by
\[
h\left(l\right)\equiv\sum_{k=1}^{l}\frac{1}{k}\,,\,\,\,\,h\left(0\right)\equiv0,
\]
and $a\left(r\right)$ is a coefficient appearing in the expansion
of $G_{DS}$ in powers of $\sin\left(\varepsilon/2\right)$ (where
$\varepsilon$ is the splitting in $\theta$), a procedure thoroughly
explained in \cite{Adam2}. Another such coefficient that will be
of use later is $d\left(r\right)$. These coefficients generally depend
on the mass of the field, on its coupling constant $\xi$, and on
the background metric. For a massless scalar field and a Schwarzschild
background geometry they assume the simple forms 
\begin{equation}
a\left(r\right)=\frac{1}{16\pi^{2}r^{2}},\,\,\,\,\,\,d\left(r\right)=-\frac{M}{24\pi^{2}r^{3}}.\label{a_c_d_Schwarz.}
\end{equation}
Therefore we have
\begin{equation}
F_{sing}\left(l,\,r\right)=-\frac{1}{2\pi r^{2}}h\left(l\right).\label{F_sing_Schwarz.}
\end{equation}
Notice that, just like $h$, $F_{sing}$ diverges logarithmically
with $l$.

We now remove the singular piece from the function $F$ by subtracting
$F_{sing}$ from it, thereby obtaining a new regularized function
denoted by $F_{reg}$ , that is
\begin{equation}
F_{reg}\left(l,\,r\right)\equiv F\left(l,\,r\right)-F_{sing}\left(l,\,r\right).\label{Freg}
\end{equation}
It turns out that subtracting $F_{sing}$ is generally insufficient
for the convergence of the sum over $l$, and that in fact, $F_{reg}\left(l,\,r\right)$
converges to a non-zero constant limit as $l\rightarrow\infty$, hence
the naive sum over $l$ of this quantity {[}multiplied by $2l+1$,
as a compensation for the denominators introduced in Eq. \eqref{coinc E 1}{]}
would diverge. This divergence reflects the fact that the counterterm
provides only partial information about the mode-sum singularity,
as information is lost in the Legendre decomposition. This obstacle
is referred to as the \emph{blind spots} phenomenon in Ref. \cite{Adam2},
and to circumvent it, a process called \emph{self-cancellation} is
employed. The idea is that the non-zero limiting value of $F_{reg}\left(l,\,r\right)$
ought to be further subtracted for the sum over $l$ to converge,
and our technical means of acheiving it is by defining the following
sequence of partial sums\footnote{Note that the limit $H\left(l\rightarrow\infty,r\right)$ is equivalent
to the sum over $l$ of the sequence $\frac{2l+1}{4\pi}\left[F_{reg}\left(l,\,r\right)-F_{reg}\left(l\rightarrow\infty,\,r\right)\right]$.}
\begin{equation}
H\left(l,\,r\right)\equiv\sum_{k=0}^{l}\frac{2k+1}{4\pi}\left[F_{reg}\left(k,\,r\right)-F_{reg}\left(l,\,r\right)\right].\label{H}
\end{equation}

The final expression for $\langle\Phi^{2}\rangle_{ren}$ is
\begin{equation}
\langle\Phi^{2}\left(x\right)\rangle_{ren}=\hbar\left[\lim_{l\rightarrow\infty}H\left(l,\,r\right)-d\left(r\right)\right].\label{phi2_ren}
\end{equation}
Recall that $H$ is constructed from functions originating in $E_{\omega l}$.
In order to obtain the Unruh state $\langle\Phi^{2}\rangle_{ren}$
from \eqref{phi2_ren}, one has to construct $H$ from $E_{\omega l}^{U}$
of Eq. \eqref{E_Unruh}. Similarly, obtaining the Hartle-Hawking state
$\langle\Phi^{2}\rangle_{ren}$ requires the use of $E_{\omega l}^{H}$
defined in \eqref{E_HH}. 

Let us summarize. The final expression for the renormalized expectation
value of $\hat{\Phi}^{2}$ is given in \eqref{phi2_ren}, with $H$
defined in \eqref{H}. There we used the function $F_{reg}$ which
is specified in \eqref{Freg}. This function involves the quantity
$F_{sing}$ defined for our specific case in \eqref{F_sing_Schwarz.}.
It further requires the use of $F$ defined in \eqref{F}, which in
turn is constructed from $E_{\omega l}$. As we mentioned above, the
latter quantity is denoted by $E_{\omega l}^{U}$ in Unruh state {[}see
\eqref{E_Unruh}{]} and by $E_{\omega l}^{H}$ in Hartle-Hawking state
{[}see \eqref{E_HH}{]}. These are computed from the radial functions,
$\bar{\psi}_{\omega l}$, defined in Eq. \eqref{psi bar} in terms
of $\psi_{\omega l}$. The latter is obtained by numerically solving
the radial equation \eqref{radial interior} with the initial conditions
\eqref{bdry cond}.

\section{Numerical implementation: $\langle\Phi^{2}\rangle_{ren}$ inside
a Schwarzschild black hole\label{numerical implementation}}

Our final expression for $\langle\Phi^{2}\rangle_{ren}$ was constructed
in Sec. \ref{PMR } using the integrands \eqref{E_Unruh} and \eqref{E_HH}.
These integrands consist of the radial function $\bar{\psi}_{\omega l}$
and also the reflection coefficient $\rho_{\omega l}^{\mathrm{up}}$
and the transmission coefficient $\tau_{\omega l}^{\mathrm{up}}$.
The computation of $\langle\hat{\Phi}^{2}\left(x\right)\rangle_{ren}$
therefore requires the numerical computation of the three quantities
$\bar{\psi}_{\omega l}\left(r\right)$, $\rho_{\omega l}^{\mathrm{up}}$
and $\tau_{\omega l}^{\mathrm{up}}$. 

The radial equation \eqref{radial interior} together with the initial
condition \eqref{bdry cond} was solved numerically for $\psi_{\omega l}$
using the ODE solver embedded in Mathematica. It was solved for 11
$l$ values ($0\leq l\leq10$), for each $l$ in the range $\omega\in\left[0,20\right]$,
with a uniform spacing $d\omega=10^{-2}$. Then $\bar{\psi}_{\omega l}$
was constructed from $\psi_{\omega l}$ using Eq. \eqref{psi bar}.

The transmission ($\tau_{\omega l}^{\mathrm{up}}$) and reflection
($\rho_{\omega l}^{\mathrm{up}}$) coefficients for each $\omega l$
mode were extracted from the numerical solution of Eq. \eqref{radial-exterior}
for the radial function $\Psi_{\omega l}$ outside the BH. Boundary
conditions were specified at the past horizon, where the radial functions
assumed the form $\Psi_{\omega l}=e^{-i\omega r_{*}}$, and the solution
was evolved towards $r\rightarrow\infty$. Here, as well, the calculation
was carried out by the ODE solver embedded in Mathematica, for the
same modes of $l$ and $\omega$ as described above. 

In order to illustrate the various stages of the renormalization procedure,
let us follow an example of the computation of $\langle\Phi^{2}\rangle_{ren}$
for $r=1.4M$ in the Hartle-Hawking state. 
\begin{figure}[h]
\subfloat[\label{E figs a}]{\includegraphics[scale=0.45]{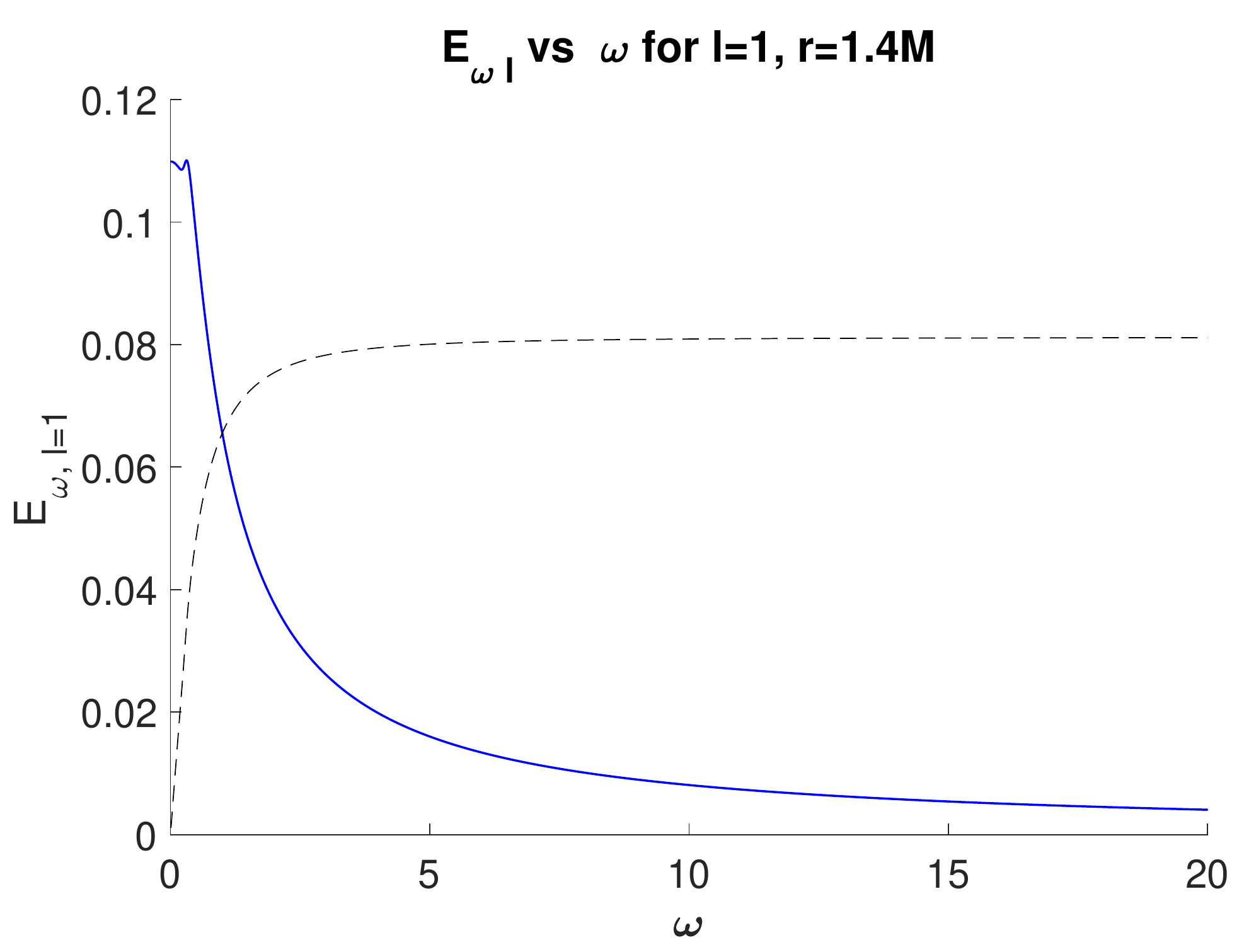}

}\subfloat[\label{E figs b}]{\includegraphics[scale=0.45]{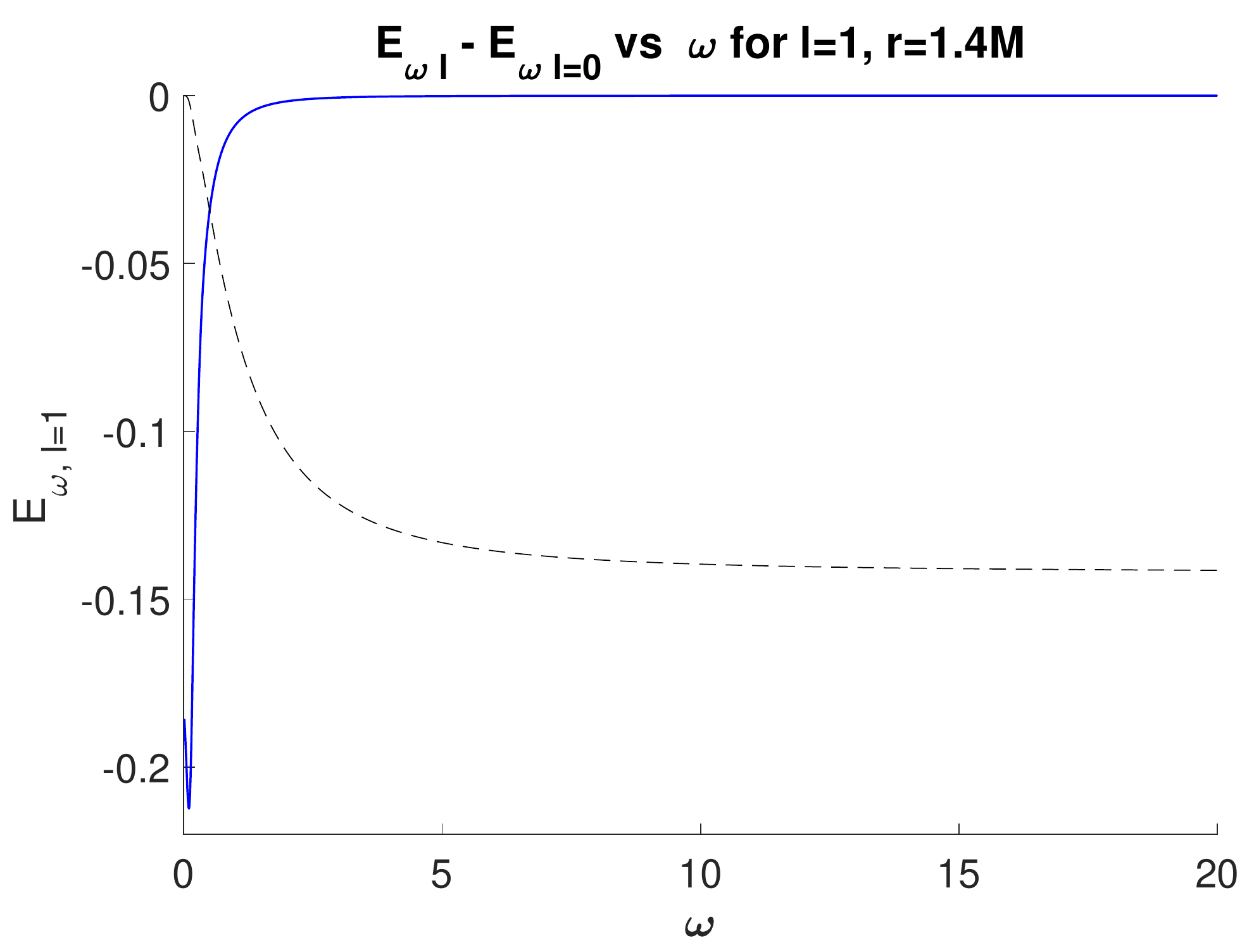}}\caption{(a) Blue curve: the numerically computed $E_{\omega,l=1}^{H}$, as
defined in \eqref{E_HH}, evaluated at $r=1.4M$. The dashed black
curve is $\omega E_{\omega,l=1}^{H}$, indicating that the asymptotic
behavior of $E_{\omega,l=1}^{H}$ at large $\omega$ is proportional
to $\omega^{-1}$. (b) Blue curve: the numerically computed difference
$E_{\omega,l=1}^{H}-E_{\omega,l=0}^{H}$, evaluated at $r=1.4M$.
The dashed black curve is $8\omega^{3}\left(E_{\omega,l=1}^{H}-E_{\omega,l=0}^{H}\right)$,
thus the asymptotic behavior of the difference at large $\omega$
is proportional to $\omega^{-3}$. \label{E figs}}
\end{figure}
 Figure \ref{E figs a} displays $E_{\omega l}^{H}$ as a function
of $\omega$ for $l=1$. Here and in all the graphs below we use units
where $M=1$ and $G=c=1$. As is indicated by the dashed black curve,
$E_{\omega l}^{H}$ behaves asymptotically at large $\omega$ as $1/\omega$,
therefore its integral diverges at infinity. This divergence is regularized
by subtracting from the integrand the $l=0$ mode, i.e. $E_{\omega,l=0}^{H}$,
as was done in Eq. \eqref{F}. This regularization results in an integrand
which behaves asymptotically as $1/\omega^{3}$, as indicated by the
dashed black curve in Fig. \ref{E figs b}, leading to a convergent
integral.

Even after the aformentioned regularization of the integral over $\omega$,
the convergence is still rather slow, and the integration requires
a very large range of $\omega$ in order to achieve a result with
sufficient accuracy. We circumvented this difficulty by employing
a large-$\omega$ expansion \footnote{We followed here the large-$\omega$ expansion procedure presented
in Appendix D in \cite{Adam2}.} of $\left|\psi_{\omega l}\left(r\right)\right|^{2}$ up to order
$\omega^{-13}$ in the integral from $\omega=20$ to infinity. 

\begin{figure}[h]
\subfloat[\label{F figs a}]{\includegraphics[scale=0.45]{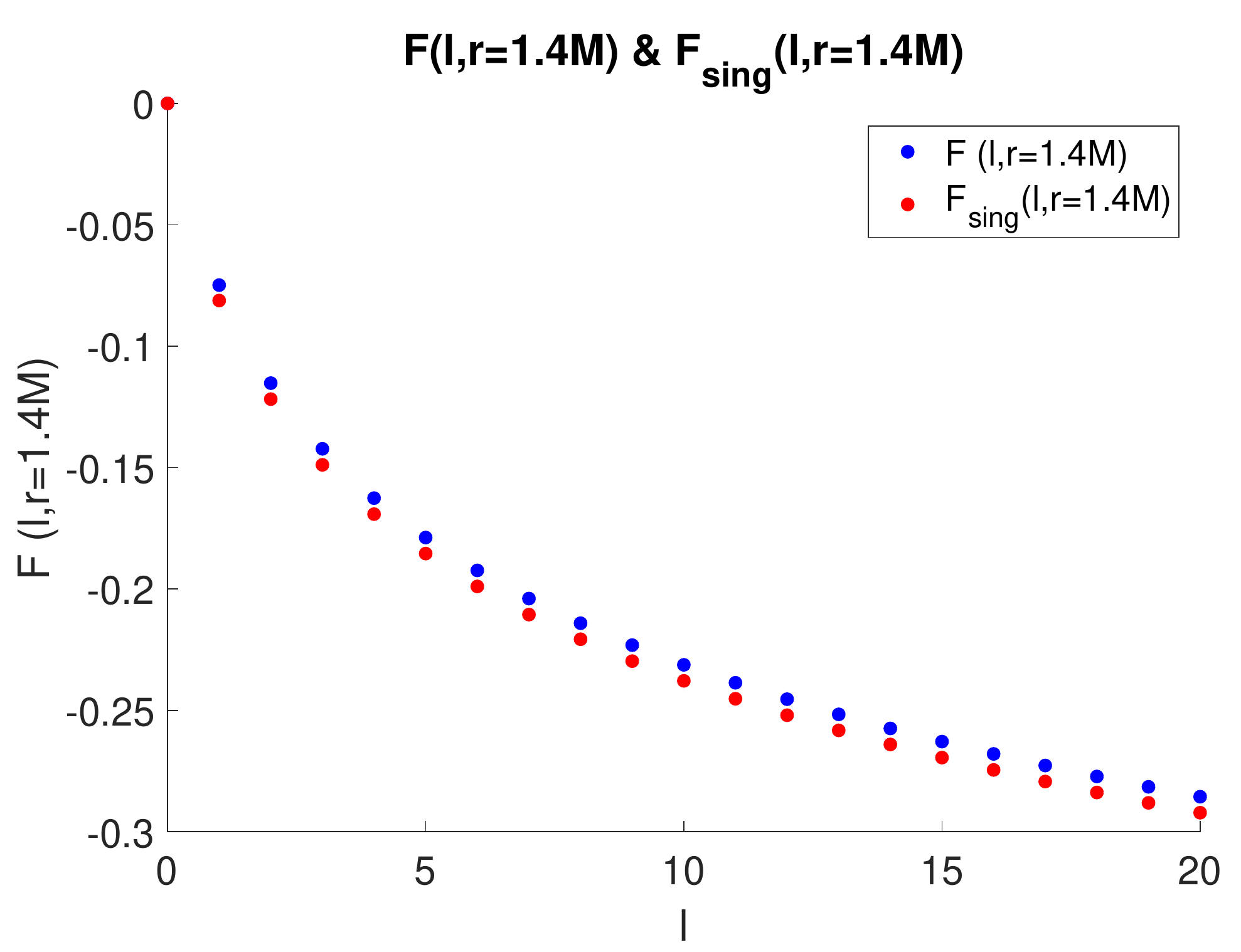}

}\subfloat[\label{F figs b}]{\includegraphics[scale=0.45]{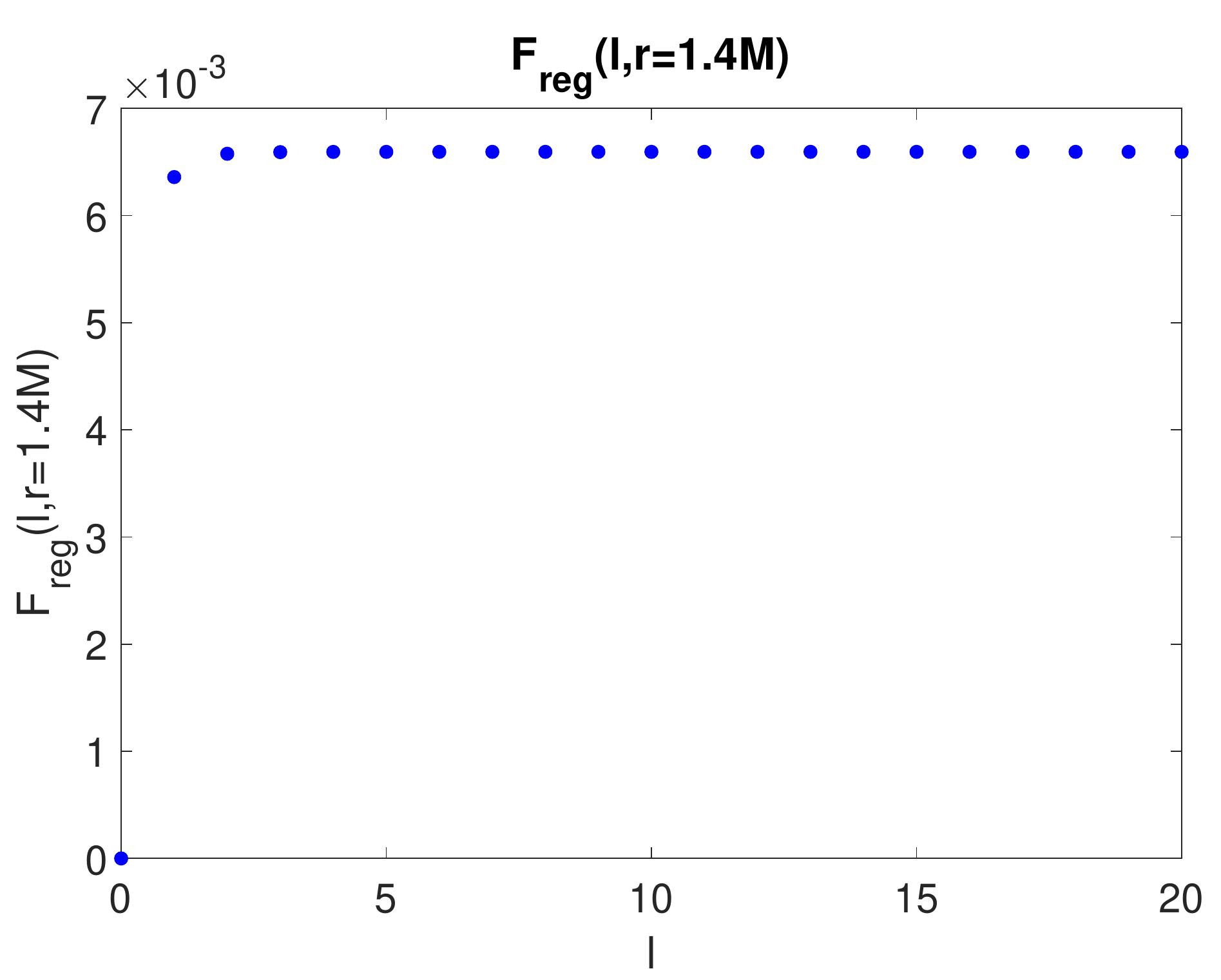}}\caption{(a) The numerically computed $F\left(l,\,r=1.4M\right)$, as defined
in \eqref{F}, represented by blue dots. The analytically computed
function $F_{sing}\left(l,\,r=1.4M\right)$, as defined in \eqref{F_sing_Schwarz.},
appears in red dots. (b) $F_{reg}\left(l,\,r=1.4M\right)$ as defined
in \eqref{Freg}, i.e. the difference between the two curves in Fig.
\ref{F figs a}. It quickly converges to a constant (the so-called
``blind-spot'' mentioned in Sec. \ref{PMR }), which will require
self-cancellation. \label{F figs}}
\end{figure}

Recall from \eqref{F} that the integral over $\omega$ gives $F\left(l,\,r\right)$,
displayed in Fig. \ref{F figs a} as a series of blue dots. 
 Note the logarithmic behavior at large values of $l$, which importantly,
characterizes $F_{sing}\left(l,\,r\right)$ as well {[}also shown
in Fig. \ref{F figs a} in red{]}. Of course, the sum over $l$ of
$\left(2l+1\right)F\left(l,\,r\right)$ then diverges, demonstrating
the need for a subtraction of the divergent piece. This is predominantly
achieved in Eq. \eqref{Freg} by subtracting $F_{sing}\left(l,\,r\right)$,
resulting in $F_{reg}\left(l,\,r\right)$ which behaves as a constant
at large values of $l$ (see Fig. \ref{F figs b}).

\begin{figure}[h]
\subfloat[\label{H figs a}]{\includegraphics[scale=0.45]{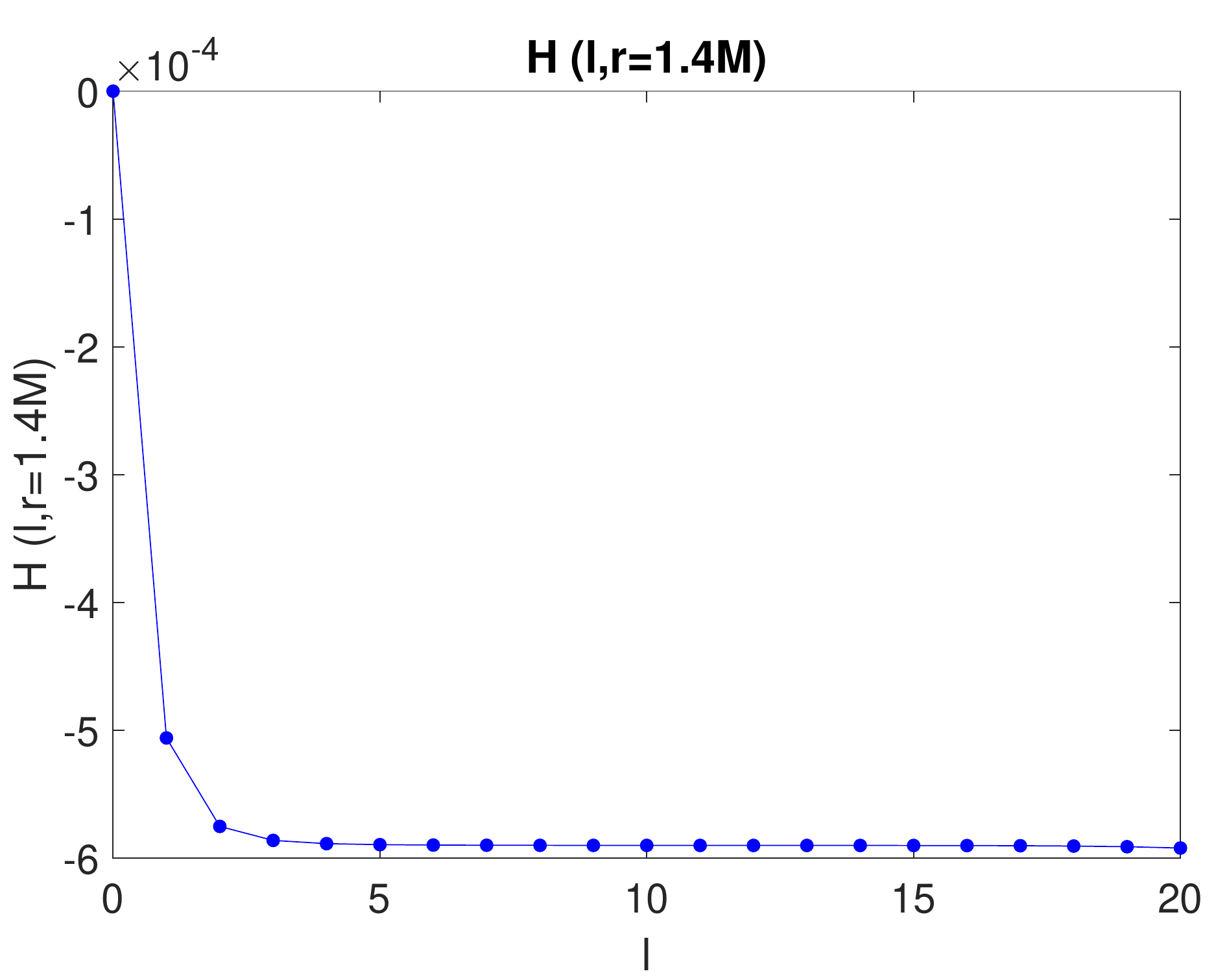}

}\subfloat[\label{H figs b}]{\includegraphics[scale=0.45]{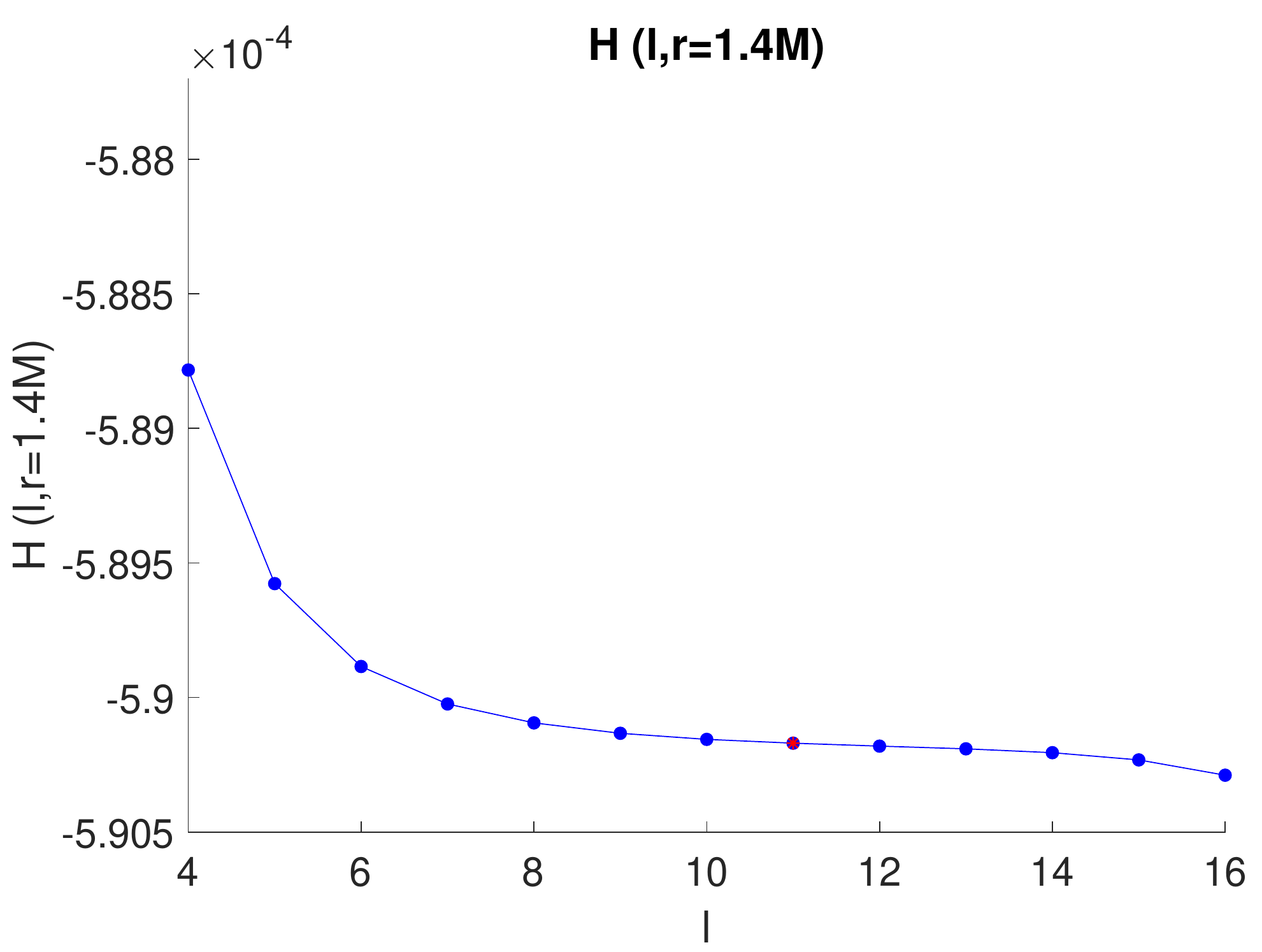}}\caption{(a) The numerically computed sequence $H\left(l,\,r=1.4M\right)$,
constructed according to Eq. \eqref{H}. It clearly rapidly converges
at large $l$. (b) A closer look at the plateau region in Fig. \ref{H figs a}.
In this scale, the growth of a numerical error at large $l$ is apparent.
The red dot indicates the estimated optimal $l$ value for the numerical
evaluation of the large-$l$ limit of $H$. This $l$ is automatically
selected by an algorithm that locates the value of $l$ beyond which
the numerical error begins to increase. \label{H figs}}
\end{figure}

We next self-cancel the ``blind spot'' by constructing the sequence
$H\left(l,\,r\right)$ according to Eq. \eqref{H}. This quantity
is presented in Fig. \ref{H figs a}. 
 Note the rapid convergence of $H\left(l,\,r\right)$, most clearly
seen in the zoom displayed in Fig. \ref{H figs b}. 

Finally, the large-$l$ limit of $H\left(l,\,r\right)$ is taken,
yielding the limiting value $-0.0005902\,M^{-2}$. When substituted
in Eq. \eqref{phi2_ren}, it yields $\langle\hat{\Phi}^{2}\left(r=1.4M\right)\rangle{}_{ren}\cong0.0009484\,\hbar M^{-2}$
in Hartle-Hawking state.

\begin{figure}
\centering{}\includegraphics[scale=0.5]{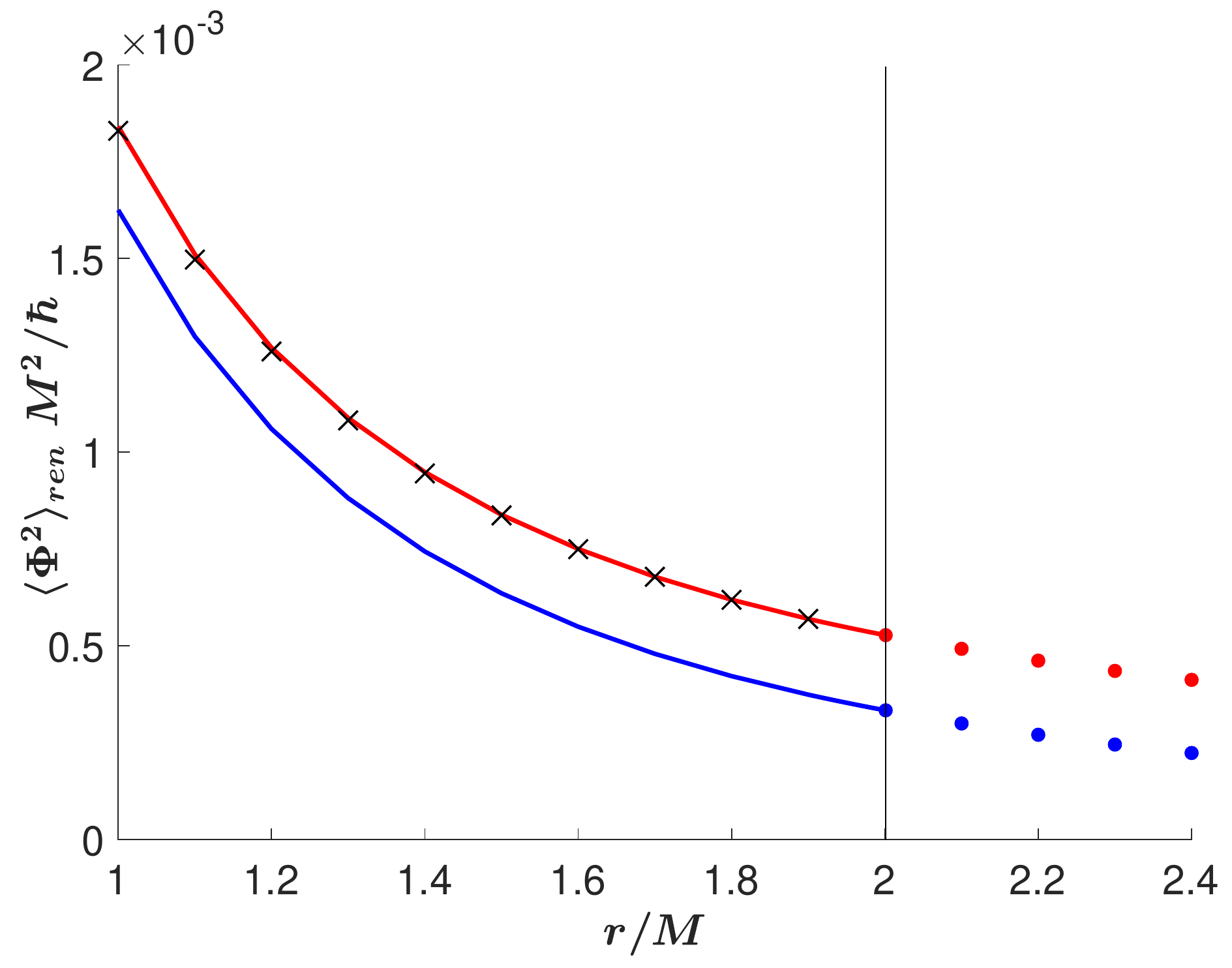}\caption{The solid red and blue lines represent our results for $<\hat{\Phi}^{2}\left(r\right)>_{ren}$
in Hartle-Hawking state and in Unruh state, respectively, inside the
BH. The red and blue dots represent the analogous results previously
obtained outside the event horizon \cite{Adam2}. The event horizon
is marked by the vertical black solid line. The results by Candelas
\& Jensen for the Hartle-Hawking state appear as black ``x'' markers.
Our results seem to agree at the event horizon with those of \cite{Adam2},
and are in a fairly good agreement with those obtained by Candelas
\& Jensen.\label{Phi2 fig}}
\end{figure}

In the same way, we computed $\langle\Phi^{2}\left(r\right)\rangle_{ren}$
for different values of $r$ in the domain\footnote{We stopped our computation at $r=1$, since the numerical difficulties
grow significantly as $r$ decreases.} $1\leq r\leq1.99999$. The same computation scheme was then applied
to the Unruh state as well. The results are presented in Fig. \ref{Phi2 fig}
which displays $\langle\Phi^{2}\rangle_{ren}$ in both Hartle-Hawking
and Unruh states for various values of $r$, together with cooresponding
results previously obtained outside the BH \cite{Adam2}. We estimate
that the numerical error is typically around two parts in $10^{3}$
or smaller. Our results for Hartle-Hawking state are compared to previous
results published by Candelas and Jensen \cite{Candelas_and_Jensen}
(black crosses), showing a fairly good agreement \footnote{With a difference of less than 1\% (which probably results from numerical
inaccuracy), improving markedly to less than 0.1\% in the region $1.6\leq r\leq1.9$.}. 

\begin{figure}[h]
\noindent \centering{}\includegraphics[scale=0.5]{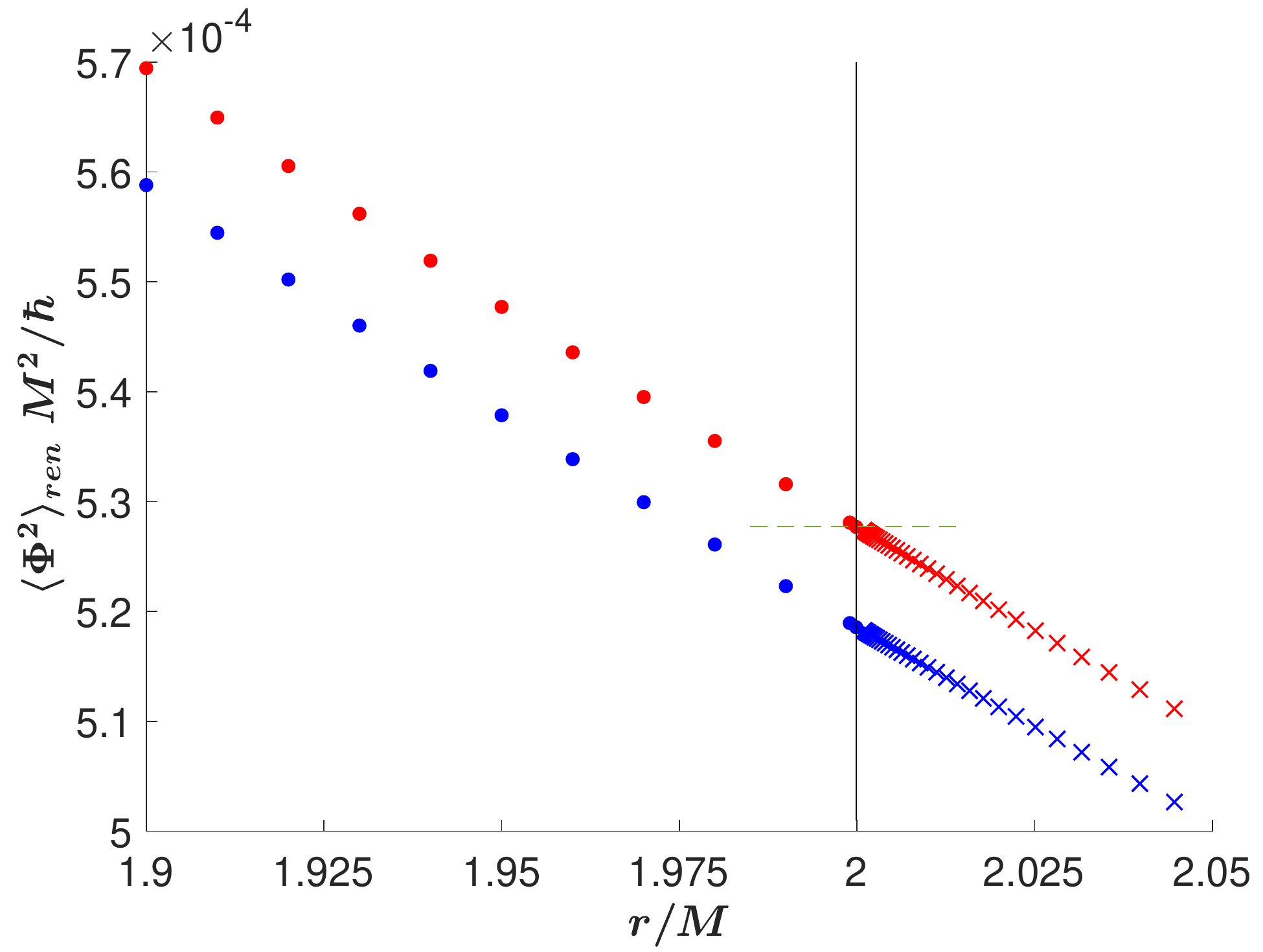}\caption{Near-horizon results for $\langle\Phi^{2}\left(r\right)\rangle_{ren}$
caclculated for different values of $r$. The red and blue dots represent
the results for $\langle\Phi^{2}\rangle_{ren}$ in Hartle-Hawking
state and in Unruh state, respectively, inside the BH. Note that the
results for the Unruh state (blue) were shifted here by an amount
$+0.000185$ for convenience of display. {[}This shift allows us to
present the two sets of data (Unruh and Hartle-Hawking) on the same
graph, with sufficient resolution at the vertical scale.{]} The red
and blue ``x'' markers represent the analogous results previously
obtained outside the event horizon, using the method described in
\cite{Adam2}. The two sets of results (external and internal ones)
nicely agree at the event horizon (vertical black solid line). They
also show good agreement with the analytical result obtained by Candelas
at the event horizon (in the Hartle-Hawking state), marked here by
the short green horizontal dashed line. \label{Phi2 Near EH}}
\end{figure}

Fig. \ref{Phi2 Near EH} displays $\langle\Phi^{2}\left(r\right)\rangle_{ren}$
for different values of $r$ in the vicinity of the event horizon,
showing results for both the interior and the exterior regions. Extrapolation
of the results, from both inside and outside the BH, to the event
horizon shows good agreement, with a difference of about $0.03\%$
(fully consistent with our estimated numerical errors). It further
shows a difference of only $\sim0.02\%$ from the analytical result
Candels obtained \cite{Candelas} for the Hartle-Hawking state $\langle\Phi^{2}\rangle_{ren}$
at the event horizon, which equals to $1/192\pi^{2}M^{2}\simeq0.0005277M^{-2}$.

\section{Discussion\label{Discussion}}

In this work we considered a massless scalar field, and numerically
computed $\langle\Phi^{2}\rangle_{ren}$ inside a Schwarzschild BH,
in both Hartle-Hawking and Unruh quantum states. We performed the
computation by employing a recently developed numerical implementation
method of the point-splitting renormalization scheme \cite{Adam1,Adam2},
which we referred to as the ``PMR'' method. This method has been
previously utilized to numerically compute $\langle\Phi^{2}\rangle_{ren}$
and $\langle\hat{T}_{\mu\nu}\rangle_{ren}$ outside Schwarzschild,
RN (unpublished) and Kerr BHs \cite{Adam1,Adam2,Adam3,Adam4,Adam5}.
Here we report the first application of the PMR method in BH interiors,
as part of an ongoing program to compute the RSET inside BHs and,
particularly, explore how it modifies the geometrical structure of
the inner (Cauchy) horizon of RN and Kerr BHs. 

In order to facilitate the computation of $\langle\Phi^{2}\rangle_{ren}$,
we used the results of Ref. \cite{group paper TPF} to express the
Hartle-Hawking and Unruh states Hadamard functions in the interior
of a Schwarzschild BH in terms of a sum of what we call \emph{inner
Eddington-Finkelstein modes}. These modes can be decomposed into radial
functions satisfying an ODE, which we solve numerically. We subsequently
employed the angular-splitting variant \cite{Adam2} of the PMR method,
and obtained $\langle\Phi^{2}\rangle_{ren}$ for Hartle-Hawking and
Unruh states inside a Schwarzschild BH. 

Our results for the Hartle-Hawking state were compared with previous
results by Candelas and Jensen \cite{Candelas_and_Jensen} with a
fairly good agreement, as seen in Fig. \ref{Phi2 fig}. The results
we obtained for both Hartle-Hawking and Unruh states further agree
at the event horizon with previous results for $\langle\Phi^{2}\rangle_{ren}$
obtained outside the BH in Ref. \cite{Adam2}. Our results for Hartle-Hawking
state at the event horizon also agree with previous analytical results
obtained by Candelas \cite{Candelas}. 

We are generally interested in the internal structure of BHs and,
in particular, in the influence of quantum fluxes on the BH interior
geometry. Understanding this issue ultimately requires the investigation
of the behavior of $\langle\hat{T}_{\mu\nu}\rangle_{ren}$ in the
interior of BHs. As a first stage towards this goal, we set out to
compute $\langle\Phi^{2}\rangle_{ren}$ in BH interiors, starting
with the case of a Schwarzschild BH as elaborated in the present paper.
It would be interesting, and important in the context of the aforementioned
research program, to further extend the present work and compute $\langle\Phi^{2}\rangle_{ren}$
inside RN \cite{Orr} and Kerr BHs. Since the latter are in fact the
real astrophysical BHs, it would be of special interest to extend
this work to Kerr BHs in the future. To accomplish that, one would
have to resort to the $t$-splitting variant (or perhaps $\varphi$-splitting
variant) of the PMR method, because the angular-splitting variant
will not work in the Kerr case. Employing the $t$-splitting variant
will have the additional benefit of providing independent results
which would serve to corroborate those presented here for the interior
of Schwarzschild BHs, obtained using the angular-splitting variant.

In the next stage, it will be necessary to extend the present analysis
of quantum effects in BH interiors from $\langle\Phi^{2}\rangle_{ren}$
to  the RSET. Especially in BHs with inner horizons, and most importantly
inside Kerr BHs. 

Beyond the challenging task of computing $\langle\hat{T}_{\mu\nu}\rangle_{ren}$
throughout the interior of BHs, in order to achieve a more complete
understanding of semiclassical BH interiors, we shall have to confront
the much harder challenge of self-consistently analyzing the back-reaction
effect of these quantum fluxes on the internal geometry (via the semiclassical
Einstein equation). Although, at the moment this goal is far out of
reach.

In addition, it would be particularly interesting to consider the
fluxes due to the quantum electromagnetic field (and perhaps also
due to quantized gravitational perturbations?). These are the more
realistic physical fields, and we hope their effects will be studied
in the future. 
\begin{acknowledgments}
This research was supported by the Asher Fund for Space Research at
the Technion.
\end{acknowledgments}

\end{document}